\documentclass[12pt,a4paper]{article}
\newcommand\be{\begin{equation}}
\newcommand\ee{\end{equation}}
\newcommand\bea{\begin{eqnarray}}
\newcommand\eea{\end{eqnarray}}
\newcommand\ket[1]{|#1\rangle}
\newcommand\bra[1]{\langle #1|}

\begin{document}
\title{Expanding Hermitean Operators in a Basis of Projectors on Coherent Spin States}
\author{Stefan Weigert \\
Institut de Physique, Universit\'e de Neuch\^atel\\
Rue A.-L. Breguet 1, CH-2000 Neuch\^atel, Switzerland\\
\tt stefan.weigert@iph.unine.ch}
\date{July 1999}
\maketitle
\begin{abstract}
The expectation values of a hermitean operator ${\widehat A}$ in $(2s+1)^2$ 
specific coherent states of a spin are known to determine the operator
unambiguously. As shown here, (almost) any other $(2s+1)^2$ coherent states 
also provide a basis for self-adjoint operators. This is proven
by considering the determinant of the Gram matrix associated with the 
coherent state projectors as a Hamiltonian of a fictitious {\em classical} spin 
system.      
\end{abstract}
\vspace{8mm}
State reconstruction \cite{leonhardt97} aims at parametrizing the density matrix 
$\hat \rho$ of a quantum system by the expectations of appropriately chosen 
observables, the quorum. For a spin $s$, the (unnormalized) density matrix has $N_s =(2s+1)^2$ 
independent real parameters; in \cite{amiet+99/2}, a particularly simple and non-redundant 
quorum consisting of precisely $N_s$ projectors on coherent spin states $\ket{\bf n}$, satisfying ${\bf n} \cdot {\widehat {\bf S}} \ket{{\bf n}} = \hbar s \ket{{\bf n}}$, has been identified. 

Indeed, the density matrix ${\hat \rho}$ of a spin $s$ is determined unambiguously 
if one performs appropriate measurements with a traditional Stern-Gerlach apparatus. 
Distribute $N_s$ axes ${\bf n}_{n}, n= 1, \ldots, N_s $, over 
$(2s+1)$ cones about the $z$ axis with different opening angles in such a way that the set 
of the $(2s+1)$ directions on each cone is invariant under a rotation about $z$ by 
an angle $2\pi/(2s+1)$. Then, an (unnormalized) statistical operator ${\hat \rho}$ is 
fixed by measuring the $(2s+1)^2 $ relative frequencies $p_s({\bf n}_n) = 
\bra{ {\bf n}_n} \hat \rho \ket{ {\bf n}_n}$, that is, by the expectation 
values of the statistical operator $\hat \rho$ in the coherent states $\ket{{\bf n}_n}$. 
In other words, a hermitean operator ${\widehat A} \in {\cal A}_s$ (which is the space of linear operators acting in the Hilbert space ${\cal H}_s$ of the spin) is fixed by the values 
of its $Q$-symbol, $Q_A({\bf n}) = \mbox{Tr} [ {\widehat A} \ket{{\bf n}}\bra{{\bf n}} ]
= \bra{{\bf n}}{\widehat A} \ket{{\bf n}}$ at $N_s$ appropriately chosen points. 
For brevity, let us denote a set of $N_s$ points (as well as the associated 
family of $N_s$ unit vectors ${\bf n}_{n}$) as a `constellation' $\cal N$ or a `hedgehog' 
$\cal N$ with unit spikes ${\bf n}_{n}$. Independent reconstruction schemes for spin $s$ do exist \cite{manko+97,agarwal98}.
    
For technical reasons, the spatial directions ${\bf n}_{n}$ dealt with in \cite{amiet+99/2} were restricted to a certain class of {\em regular} hedgehogs, ${\cal N}_0$. The purpose here is to show that this restriction is not necessary: given a {\em generic} constellation $\cal M$, the $N_s$ values of the Q-symbol $Q_A({\bf n}_n)$ contain all the information about the operator $\widehat A$. Let us put it differently: given {\em any} constellation $\cal M$ of vectors ${\bf m}_n$, then {\em either} the numbers $Q_A ({\bf m}_n)$  determine $\widehat A$, {\em or} there is an {\em infinitesimally close} constellation ${\cal M}'$ such that the numbers $Q_A ({\bf m}^\prime_n)$ do the job. Two hedgehogs ${\cal M}'$ and ${\cal M}$ are close if, for example, the number 
\be 
d({\cal M}' , {\cal M}) = \sum_{n=1}^{N_s} | {\bf m}^\prime_{n} - {\bf m}_{n} | \, ,
\label{distance}
\ee
is small. To visualise this statement, consider the real vector space $I\!\!R^3$: any three unit vectors form a basis provided they are neither co-planar nor co-linear. Among all possibilites, the exceptional constellations have measure zero. At the same time, it is obvious that arbitrarily small variations typically turn the three linearly dependent vectors into a basis of $I\!\!R^3$.   

The starting point of the proof are $N_s$ projection operators on coherent states,
\be
{\widehat Q}_n 
= \ket{{\bf n}_{n}} \bra{{\bf n}_{n}}   \, , \quad {\bf n}_n \in {\cal N}^0 \, ,  
          \qquad  1 \leq n \leq N_s  \, ,
\label{mixedrho}
\ee
determined uniquely by the constellation ${\cal N}_0$ described. It will be shown now any  other hedgehog ${\cal M}$ (or an infinitesimally close one, ${\cal M}'$) also will provide a basis of the space ${\cal A}_s$. 

The $N_s^2$ elements of the {\em Gram matrix} ${\sf G}_{nn'}$ \cite{greub63} associated with a constellation $\cal M$ are given by the scalar product of the projectors on coherent states:
\be
 {\sf G}_{nn'} = \mbox{ Tr } \left[ {\widehat Q}_{n} {\widehat Q}_{n'} \right] 
               = | \bra{{\bf m}_{n}} {\bf m}_{n'} \rangle |^2
               = \left( \frac{ 1 + {\bf m}_{n} \cdot {\bf m}_{n'}}{2} \right)^{2s} \, ,
\qquad  1 \leq n,n' \leq N_s \, .
\label{gram}
\ee
Thus, the scalar product of two coherent states is a {\em polynomial} in the components of the associated unit vectors ${\bf m}_{n}$ and  ${\bf m}_{n'}$. The result in \cite{amiet+99/2} comes down to saying that the Gram matrix of the constellation ${\cal N}_0$ is invertible or, equivalently, its determinant does not vanish.

The determinant of the matrix ${\sf G}$, if conceived as a function of the $n$-th
vector, is infinitely often differentiable with respect to its components, according to (\ref{gram}). Upon keeping the vectors ${\bf n}_1, \ldots,{\bf n}_{n-1}$ and ${\bf n}_{n+1},\ldots, {\bf n}_{N_s}$ fixed, it may be regarded as a fictitious time-independent {\em Hamiltonian function} $H$ of a single classical spin, ${\bf n}_n$:
\be 
\det {\sf G} ({\bf n}_n) = H ({\bf n}_n) \, .
\label{hamilton}
\ee
It is different from zero if ${\bf n}_n$ coincides with the $n$-th vector of the constellation
${\cal N}_0$. This Hamiltonian describes an {\em integrable} system since there is just one degree of freedom accompanied by one constant of motion, the Hamiltonian itself \cite{arnold84}. The two-dimensional phase space ${\sf S}^2$ is foliated entirely by one-dimensional tori of constant energy. In addition, a finite number of (elliptic or hyperbolic) fixed points and one-dimensional separatrices will occur. This can be seen, for example, by looking at the flow on the unit sphere generated by the Hamiltonian $H ({\bf n}_n)$:
\be 
\frac{d {\bf n}_n}{dt} = {\bf n}_n \times \frac{\partial H}{\partial {\bf n}_n} \, ,
\label{spindyn}
\ee
where $\partial / \partial {\bf n}_n$ is the gradient with respect to ${\bf n}_n$ \cite{srivastava+87}. The right-hand-side is a (non-zero) polynomial in the components of ${\bf n}_n$, implying that the integral curves of the Hamiltonian are fixed points, separatrices, and closed orbits. This means that $H ({\bf n}_n)$ can take
the value zero at a finite number of (open or closed) curves or points at most. Consequently, the determinant of ${\sf G} ({\bf n}_n)$ is different from zero for almost all choices of ${\bf n}_n$. Therefore, one can move the vector ${\bf n}_n$ into any other vector, including ${\bf m}_n$, the $n$-th vector of the desired constellation 
${\cal M}$, thereby passing possibly through points with $\det {\sf G} = 0$. If, accidentally,  ${\bf m}_n$ corresponds to a point with vanishing energy (this happens with probability zero only), one can nevertheless approach it arbitrarily close by a vector ${\bf m}^\prime_n$ with $| {\bf m}^\prime_n - {\bf m}_n | < \varepsilon /N_s$ since levels of constant energy have a co-dimension at most equal to one. 

Working one's way from $n=1$ to $N_s$, one ends up with a constellation ${\cal M}'$ which is guaranteed to be infinitesimally close to ${\cal M}$ since $\sum_n | {\bf m}^\prime_n - {\bf m}_n | < \varepsilon$ can be made arbitrarily small. With probability one, the constellation $\cal M$ is obtained even exactly. Consequently, almost all hedgehogs ${\cal M}$ of $N_s$ projection operators ${\widehat Q}_{n}$ give rise to a {\em basis} in the space of linear operators on ${\cal H}_s$, the Hilbert space of a spin $s$. In turn, the values of the {\em discrete} Q-symbol related to a constellation $\cal M$ are indeed sufficient to determine the operator $\widehat A$.  

In summary, it has been shown that (almost) any distribution of $N_s$ points on the sphere ${\sf S}^2$ gives rise to a non-orthogonal basis of coherent-state projectors ${\widehat Q}_n$ in the linear space ${\cal A}_s$ of operators for a spin $s$. An independent proof of this result can be found in \cite{amiet+99/3}. In addition, a discrete variant of the $P$-symbol is shown there to come along naturally with the discrete $Q$-symbol. The relation of the basis 
of projectors ${\widehat Q}_n$ to a symbolic calculus {\em \`a la} Stratonovich-Weyl has been elaborated in \cite{weigert99/2}. 
\subsection*{Acknowledgements}
The author acknowledges financial support by the {\em Schweizerische Nationalfonds}.  

\end{document}